%% file: main.tex
\input{macros}

\begin{document}

\title{\name: Fuzzing Sparse Tensor Compilers (Registered Report)}
% \author{Anonymous Authors}
\input{authors/authors}

\IEEEoverridecommandlockouts

% Conference header block
\makeatletter\def\@IEEEpubidpullup{6.5\baselineskip}\makeatother
\IEEEpubid{\parbox{\columnwidth}{
		Fuzzing Workshop (FUZZING) 2026\\
		27 February 2026, San Diego, CA, USA\\
		ISBN 978-1-970672-04-6\\  
		https://dx.doi.org/10.14722/fuzzing.2026.23006\\
		www.ndss-symposium.org
}
\hspace{\columnsep}\makebox[\columnwidth]{}}

\maketitle

\input{sections/abstract}

\input{sections/introduction}
\input{sections/background}
\input{sections/motivation}

\input{sections/design_implementation}
\input{sections/evaluation}
\input{sections/discussion}
\input{sections/related_work}

\input{sections/conclusion}

\bibliographystyle{IEEEtran}
\bibliography{ref}

\end{document}

%% file: macros.tex
\documentclass[conference]{IEEEtran}
\usepackage{cite}
\usepackage[hidelinks]{hyperref}
\usepackage{amsmath,amssymb,amsfonts}
\usepackage{graphicx}
\usepackage{textcomp}
\usepackage{color}
\usepackage{adjustbox}
\usepackage{makecell}
\usepackage{multirow}
\usepackage{threeparttable}
\usepackage{enumitem}
\usepackage{booktabs}
\usepackage{bbding}
\usepackage{colortbl}
\usepackage{xcolor}
\usepackage{setspace}
\usepackage{pifont}
\usepackage{caption}
\usepackage{subcaption}
\usepackage{fontawesome}
\usepackage{listings}
\usepackage[frozencache,cachedir=.]{minted}
\usepackage[ruled,linesnumbered,vlined]{algorithm2e}
\usepackage[noend]{algpseudocode}
\usepackage[most]{tcolorbox}
\usepackage{float}
\newcounter{matrixctr}
\renewcommand{\thematrixctr}{M\arabic{matrixctr}}

\setminted{
    frame=lines,      % Adds a frame at top and bottom
    framesep=2mm,     % Distance between frame and code
    baselinestretch=1.2,
    fontsize=\footnotesize, % Essential for 2-column papers to fit code
    linenos,          % Adds line numbers
    breaklines        % wrap long lines (CRITICAL for 2-column)
}

\newcommand{\eg}{\textit{e.g.,}~}
\newcommand{\ie}{\textit{i.e.,}~}

\newcommand{\name}{\textsc{TenSure}\xspace}

%% file: authors/authors.tex
% author names and affiliations
% use a multiple column layout for up to three different
% affiliations
% \author{\IEEEauthorblockN{Kabilan Mahathevan}
% 	\IEEEauthorblockA{Virginia Tech\\
% 		kabilan@vt.edu}
% 	\and
% 	\IEEEauthorblockN{Kabilan Mahathevan}
% 	\IEEEauthorblockA{Virginia Tech\\
% 		kabilan@vt.edu}
% 	\and
% 	\IEEEauthorblockN{Kabilan Mahathevan}
% 	\IEEEauthorblockA{Virginia Tech\\
% 		kabilan@vt.edu}
%     \and
%     \IEEEauthorblockN{Kabilan Mahathevan}
% 	\IEEEauthorblockA{Virginia Tech\\
% 		kabilan@vt.edu}}
	
% conference papers do not typically use \thanks and this command
% is locked out in conference mode. If really needed, such as for
% the acknowledgment of grants, issue a \IEEEoverridecommandlockouts
% after \documentclass

% for over three affiliations, or if they all won't fit within the width
% of the page, use this alternative format:
% 
\author{
\IEEEauthorblockN{Kabilan Mahathevan, Yining Zhang, Muhammad Ali Gulzar, Kirshanthan Sundararajah}
\IEEEauthorblockA{Department of Computer Science\\
Virginia Tech, Blacksburg, VA 24061, USA\\
Emails: \{kabilan, yiningz, gulzar, kirshanthans\}@vt.edu}
}

%% file: sections/abstract.tex
\begin{abstract}

Sparse Tensor Compilers (STCs) have emerged as critical infrastructure for optimizing high-dimensional data analytics and machine learning workloads. 
The STCs must synthesize complex, irregular control flow for various compressed storage formats directly from high-level declarative specifications, thereby making them highly susceptible to subtle correctness defects. 
Existing testing frameworks, which rely on mutating computation graphs restricted to a standard vocabulary of operators, fail to exercise the arbitrary loop synthesis capabilities of these compilers. 
Furthermore, generic grammar-based fuzzers struggle to generate valid inputs due to the strict rules governing how indices are reused across multiple tensors.

In this paper, we present \name, the first extensible black-box fuzzing framework specifically designed for the testing of STCs. 
\name leverages Einstein Summation (Einsum) notation as a general input abstraction, enabling the generation of complex, unconventional tensor contractions that expose corner cases in the code-generation phases of STCs. 
We propose a novel constraint-based generation algorithm that guarantees 100\% semantic validity of synthesized kernels, significantly outperforming the $\sim$3.3\% validity rate of baseline grammar fuzzers. 
To enable metamorphic testing without a trusted reference, we introduce a set of semantic-preserving mutation operators that exploit algebraic commutativity and heterogeneity in storage formats. 
Our evaluation on two state-of-the-art systems, TACO and Finch, reveals widespread fragility, particularly in TACO, where \name exposed crashes or silent miscompilations in a majority of generated test cases. 
These findings underscore the critical need for specialized testing tools in the sparse compilation ecosystem.

\end{abstract}

%% file: sections/introduction.tex
\section{Introduction} \label{sec:introduction}

Over the past decade, the demand for efficient execution of scientific computing and deep learning workloads has surged, pushing tensor compilers to the centre of modern systems. 
Compiler frameworks such as TVM~\cite{tvm}, TensorRT~\cite{tensor_rt}, Triton~\cite{triton}, and XLA~\cite{xla} aggressively optimise tensor computations by lowering high-level programs into hardware-specific code.
Concurrently, tensor sizes are expanding rapidly while hardware scaling plateaus~\cite{endofmoore}. 
This divergence necessitates performing computation directly on compressed tensor storage formats.

However, it is non-trivial to construct optimized tensor computations, or kernels, for compound operations on various compressed tensor data storage formats. 
Unlike dense arrays, compressed formats do not support efficient random access; consequently, the resulting kernels require sophisticated loop nests to synchronize access across multiple sparse operands. 
Therefore, a compiler-based approach is widely advocated for sparse tensor operations.
This need has driven the emergence of Sparse Tensor Compilers (STCs) as a dedicated class of tools for performing computations on compressed storage formats~\cite{taco, Ahrens2025Finch, sparse_tir, sparse_mlir, aart_bik_spmv, pytorch_sparse}.

Unfortunately, the complexity of sparse traversal schedules significantly exceeds that of dense counterparts~\cite{auto_spmv_gpu, taco, sparse_mlir, aart_bik_spmv}.
The need to support diverse compressed data structures creates a large surface area for subtle correctness bugs\textemdash including iteration faults and memory safety violations\textemdash particularly when chaining multiple tensor operations.

Recent real-world failures demonstrate that compiler bugs can cause severe damage even in simpler dense settings.
For instance, Anthropic traced a degradation in Claude’s responses to an XLA miscompilation~\cite{anthropic_bug}.
Diagnosing such issues is notoriously difficult because model outputs reveal little about the underlying compiled program, and developers implicitly trust the compiler.
Sparse tensor pipelines amplify this challenge: the control flow is far more intricate, making defects significantly harder to isolate.

In this paper, we introduce an extensible automated testing framework specifically designed to verify the functional correctness of STCs. 
Existing approaches based on differential testing \cite{csmith, yarpgen} to validate general-purpose compilers are fundamentally incompatible with tensor compilers because of the non-negligible number of false positives due to variations in numerical accuracy across different execution environments \cite{metamorphic_dnn}.
Therefore, \name adopts an approach similar to \textsc{PolyJuice}, leveraging the same execution environment for metamorphic testing.

Conventional semantic-preserving mutation techniques used in traditional metamorphic testing, such as dead-code injection~\cite{compiler_emi}, instruction padding~\cite{glfuzz}, and control-flow restructuring~\cite{csmith, clsmith}\textemdash rely on mutating imperative control structures (\eg if blocks, loops, and statement ordering).
However, STCs typically operate on high-level declarative specifications, such as {\em Einsum} (Einstein summation) notation~\cite{einstein1916foundation}, where explicit control flow is absent. 
Consequently, these standard imperative mutations are inapplicable, as the control flow is not defined in the source but is synthesized during compilation, leaving no target for traditional code transformations.

Furthermore, the landscape of STCs is characterized by significant syntactic and semantic fragmentation. 
Prominent frameworks like TACO~\cite{taco}, Finch~\cite{Ahrens2025Finch}, and the MLIR Sparse Dialect~\cite{sparse_mlir} each expose distinct declarative DSLs and support varying subsets of the einsum standard. 
This divergence is compounded by heterogeneous underlying infrastructures\textemdash ranging from C++ template libraries to LLVM-based lowering pipelines. 
Therefore, creating a unified testing framework is non-trivial; a test case valid for one tool is syntactically incompatible with another. 
To bridge this interoperability gap, an intermediate representation is required to abstract the core semantic features of STC programs, decoupling test generation from implementation-specific syntax.

Previous work on testing dense tensor compilers, such as NNSmith~\cite{nnsmith} and PolyJuice~\cite{polyjuice}, relies on constructing computation graphs\textemdash directed acyclic structures where nodes represent fixed high-level operators (\eg MatMul, Conv2d) connected by data dependencies. 
This approach is inherently limited by the fixed vocabulary of the standard operator library; it tests the compiler's ability to optimize known patterns but fails to stress-test its capability to synthesize loop nests for arbitrary, unconventional tensor operations.
Crucially, these frameworks also lack any representation of compressed storage formats, the backbone of STCs, thereby missing the complex traversal constraints required for sparse compilation.

To the best of our knowledge, no existing testing framework specifically targets functional correctness for STCs. This leaves a substantial portion of the tensor-compiler ecosystem effectively untested.
We address this gap by developing a fuzzer that generates valid tensor kernels, translates them into the target sparse-tensor DSL, and creates semantically equivalent program variants for metamorphic testing. 
Unlike general-purpose fuzzing, where large sets of rewrite rules yield many equivalent variants, single-kernel tensor programs offer limited opportunities for mutation.

% Guided by these constraints, our fuzzer employs two mutation strategies\textemdash detailed in Section \ref{sec:design}\textemdash to verify functional correctness. 
% First, we leverage \textit{Algebraic Commutativity} by reordering operands within the Einsum equation. This tests whether the compiler respects algebraic properties and generates valid code for different operand permutations. 
% Second, we exploit \textit{Storage Format Heterogeneity}. 
% Since the logical result of a tensor contraction is independent of its physical representation, we vary the compressed formats of all input and output tensors. 
% By enforcing that the computation yields deterministic results regardless of the underlying data layout, we effectively stress-test the compiler's ability to handle diverse sparse iteration patterns.

We evaluated \name on two state-of-the-art systems: the C++-based TACO~\cite{taco} and the Julia-based Finch~\cite{Ahrens2025Finch}. 
Our experiments revealed significant robustness issues in TACO, where the fuzzer exposed crash-inducing inputs or miscompilations in over 60\% of generated test cases. 
Furthermore, the successful integration with Finch validates the framework's extensibility to diverse compiler architectures. 
Collectively, these preliminary results highlight the pervasive fragility of the current sparse tensor compilation infrastructure and demonstrate the efficacy of \name in detecting latent defects.

This paper makes the following contributions:
\begin{itemize}
    \item \textbf{Extensible STC Fuzzer}: To the best of our knowledge, we present \name, the first language-agnostic black-box fuzzing framework specifically designed to validate STCs.
    \item \textbf{Domain-Specific Mutation Operators}: We define a set of mutation operators that exploit \textit{Storage Format Heterogeneity} and \textit{Algebraic Commutativity} to generate semantically equivalent STC programs.
    \item \textbf{Constraint-Based Generation Algorithm}: We formalize and implement a generation algorithm that solves context-sensitive dimensional constraints to synthesize einsum expressions. Unlike standard grammar-based fuzzers, which achieve a validity rate of only $\sim$3.3\%, our approach guarantees 100\% semantic validity, enabling high-throughput testing of deep compilation passes.
\end{itemize}

The remainder of this paper is structured as follows. 
Section \ref{sec:background} provides background on sparse compilation, while Section \ref{sec:motivation} motivates the need for specialized fuzzing.
The details of the design of \name are provided in the Section \ref{sec:design}.
We present a preliminary evaluation of the tool in Section \ref{sec:evaluation}
Sections \ref{sec:discussion} and \ref{sec:related} discuss the limitations and related work, respectively, and we conclude the paper in Section \ref{sec:conclusion}.

%% file: sections/background.tex
\section{Background}\label{sec:background}

\subsection{Tensors and Einsum}

Tensors are multi-dimensional arrays, and their computations can be concisely expressed using einsum notation~\cite{RicciLeviCivita1901}. 
The einsum notation generalizes tensor operations by implying summation over shared indices between tensors, eliminating the need for explicit summation symbols. 

%Equation~\ref{eq:bug-einsum} shows a simple kernel with tensor contraction expressed in einsum notation. 
$A(j) = B(i, j) * C(i)$ is a simple kernel with tensor contraction expressed in einsum notation.
%As formalized by the summation in Equation~\ref{eq:bug-einsum-summation}, this operation defines the output tensor $A_j$ (or $A(j)$) by accumulating the product of $B_{ij}$ and $C_i$ along the shared index $i$. 
As formalized by the summation $A_j = \smash{\sum_{i}} B_{ij} C_i$, this operation defines the output tensor $A_j$ (or $A(j)$) by accumulating the product of $B_{ij}$ and $C_i$ along the shared index $i$. 
It is important to note that while einsum specification is declarative\textemdash defining the data dependencies rather than the execution flow\textemdash it implies a \textit{reduction} over the $i$ dimension.
In other words, this operation performs a reduction over the $i$ dimension while accumulating contributions from $B_{ij}$ weighted by $C_i$. 
%For reference, Equation~\ref{eq:mm-einsum} presents the standard General Matrix–Matrix Multiplication (GeMM) using the declarative einsum notation, while Equation~\ref{eq:mm-einsum-summation} shows the corresponding algebraic definition expressed as summation.
For reference, $A(i,j) = B(i,k) * C(k,j)$ presents the standard General Matrix–Matrix Multiplication (GeMM) using the declarative einsum notation, while $A_{ij} = \smash{\sum_{k}} B_{ik} C_{kj}$ shows the corresponding algebraic definition expressed as summation.

% \noindent
% \begin{minipage}[t]{.48\linewidth}
%     \small
%     \begin{equation}
%         A({j}) = B({i,j}) * C({i}) \vphantom{\sum_{i}}
%         \label{eq:bug-einsum}
%     \end{equation}
% \end{minipage}
% \hfill
% \begin{minipage}[t]{.48\linewidth}
%     \small
%     \begin{equation}
%         A_j = \smash{\sum_{i}} B_{ij} C_i
%         \label{eq:bug-einsum-summation}
%     \end{equation}
% \end{minipage}

% % \medskip

% \noindent
% \begin{minipage}[t]{.48\linewidth}
%     \small
%     \begin{equation}
%         A({i,j}) = B({i,k}) * C({k,j}) \vphantom{\sum_{k}}
%         \label{eq:mm-einsum}
%     \end{equation}
% \end{minipage}%
% \hfill
% \begin{minipage}[t]{.48\linewidth}
%     \small
%     \begin{equation}
%         A_{ij} = \smash{\sum_{k}} B_{ik} C_{kj}
%         \label{eq:mm-einsum-summation}
%     \end{equation}
% \end{minipage}

\subsection{Tensor Compilers and Loop Lowering}

Tensor compilers, such as TVM~\cite{tvm}, XLA~\cite{xla}, and MLIR~\cite{mlir}, serve as bridges between high-level mathematical notation and efficient machine code. 
To insulate scientists from low-level implementation details, these frameworks expose declarative DSLs. 
While einsum notation serves as a powerful generic abstraction for defining arbitrary contractions, modern compilers also support direct declarative definitions for ubiquitous tensor operations, such as matrix multiplications and 2D convolutions.
% Existing testing frameworks primarily leverage these high-level definitions to construct computation graphs as input.

The fundamental task of these compilers is lowering\textemdash the automated translation of these declarative specifications into optimized, imperative loop nests. 
Algorithm~\ref{alg:bug-dense-kernel} presents a pseudo-code representation of the lowering output for the kernel $A(j) = B(i, j) * C(i)$. 
Even for dense tensors, this translation is non-trivial: a concise mathematical expression expands into a complex sequence of nested loops, explicit memory addressing, and boundary checks. 
Consequently, the cyclomatic complexity of the generated code scales rapidly with the dimensionality of the tensors and the depth of the operation chain, creating significant potential for faults during lowering.

\input{algorithms/taco-sparse-kernel}

\subsection{Compressed Data Structure}

Unlike dense tensors, where almost all entries are nonzero, most real-world large-scale tensors are highly sparse. 
For example, the Amazon Reviews tensor in particular, contains 1.5 x 10$^{19}$ components corresponding to 107 exabytes of data (assuming 8 bytes are used per component), but only 1.7 × 10$^{9}$ of the components (13 gigabytes) are non-zero~\cite{netflix_tensor, facebook_tensor, amazon_review_tensor, taco}. 
Sparse tensors address this imbalance by representing data in compressed formats such as Coordinate (COO), Compressed Sparse Row (CSR), Compressed Sparse Column (CSC)~\cite{duff2017direct}, or more specialized hybrid layouts. 
These formats store only nonzero values, along with their index metadata, allowing compilers to skip redundant iterations and avoid unnecessary multiplications and additions by zero. 
This is critical for achieving efficient computation on large-scale tensor workloads.

\subsection{Sparse Tensor Compilers}

The Tensor Algebra Compiler (TACO)~\cite{taco} pioneered the automated generation of sparse tensor kernels and serves as a primary subject of our evaluation. 
Following this precedent, major infrastructure frameworks including MLIR~\cite{sparse_mlir}, TVM~\cite{sparse_tir}, XLA~\cite{tensorflow_sparse}, and PyTorch~\cite{scorch, pytorch_sparse} have introduced sparse extensions. 
However, despite this broad interest, support for sparse operations in these systems remains largely experimental. 
The inherent complexity of sparse compilation has hindered the development of fully robust, production-ready implementations, resulting in a landscape of beta-level features that lack correctness guarantees. 
This pervasive fragility underscores the critical need for a dedicated testing framework to validate these evolving STCs.

\subsection{Automated Testing Techniques}

Compiler validation typically relies on the synergy between robust test oracles and structured input generation.
Differential and metamorphic testing has established itself as the gold standard for validating general-purpose and dense tensor compilers~\cite{csmith, compiler_emi, yarpgen, clsmith, nnsmith, polyjuice}. 
Fundamentally, this approach is black-box and relies on the invariant that semantically equivalent inputs\textemdash or the same input executed on different compiler implementations\textemdash must yield identical execution results.
This simplicity allows it to detect subtle functional correctness bugs without requiring a formal specification of the compiler's internal logic.

% In practice, this strategy is realised through two distinct methodologies. 
% The first, often termed Metamorphic Testing, generates semantically equivalent variants of a source program and verifies that the target compiler produces consistent outputs for all variants. 
% The second approach, Cross-Implementation Testing, executes a single source program across multiple distinct compiler implementations and checks for output divergence. 
% While popular in standardised languages like C, the cross-implementation approach introduces a significant oracle ambiguity: when compilers disagree, identifying the culprit is non-deterministic without a ground truth. 
% Furthermore, as noted in Section \ref{sec:introduction}, the syntactic fragmentation of STCs makes executing the exact same source program across different tools practically infeasible.

To drive these differential and metamorphic testing campaigns, compilers require highly structured inputs that satisfy strict syntactic rules.
Grammar-based Fuzzing addresses this by generating syntactically valid test cases through adherence to a formal language specification, typically defined using standard Backus-Naur Form (BNF) or tool-specific Extended BNF (EBNF) dialects such as ANTLR.
Unlike unstructured mutational fuzzers that apply random bit-flips to seed inputs, grammar-based tools, such as Grammarinator~\cite{grammarinator}, LangFuzz~\cite{langfuzz}, and Nautilus~\cite{nautilus} construct inputs by performing random walks over the grammar's derivation tree.
This technique is particularly effective for testing language processors, as it ensures that the generated inputs successfully pass the parser and exercise the deeper semantic analysis and lowering phases of the compiler.

% This approach is particularly effective for testing compilers and interpreters where the input space is highly structured, as it guarantees that generated test cases satisfy the syntactic rules of the target language. 
% However, standard grammar-based fuzzers generally operate on Context-Free Grammars (CFGs). While they excel at constructing valid syntax trees, they lack the intrinsic capability to enforce complex semantic constraints.

%% file: algorithms/taco-sparse-kernel.tex
\begin{figure*}[t] 
    % \captionsetup{justification=left}
    \noindent 
    % --- LEFT COLUMN ---
    \begin{minipage}[b]{0.48\textwidth}
        \begin{minted}[fontsize=\scriptsize, frame=lines, framesep=2mm]{c}
int main() {
  int j_A = 0;
  for (int i_C=C_col_ptr[0]; i_C < C_col_ptr[1]; i_C++) {
    int i = C_row_ind[iC];
    for (int j = 0; j < B2_dimension; j++) {
      // B2_dimension->Dimension of the B's column axis
      A_values[j_A] = 0.0;
      int j_B = i * B2_dimension + j;
      A_values[j_A] += B_values[j_B] * C_values[i_C];
      j_A++;
    }
  }
}
        \end{minted}
        \captionof{listing}{TACO Generated Program (Buggy)}
        \label{lst:taco_buggy_code}
    \end{minipage}%
    \hfill 
    % --- RIGHT COLUMN ---
    \begin{minipage}[b]{0.48\textwidth}
        \begin{minted}[fontsize=\scriptsize, frame=lines, framesep=2mm]{c}
int main() {
  for (int i_C=C_col_ptr[0]; i_C < C_col_ptr[1]; i_C++) {
    int i = C_row_ind[i_C];
    int j_A = 0;
    for (int j = 0; j < B2_dimension; j++) {
    // B2_dimension->Dimension of the B's column axis
      int j_B = i * B2_dimension + j;
      A_values[j_A] += B_values[j_B] * C_values[i_C];
      j_A++;
    }
  }
}
        \end{minted}
        \captionof{listing}{Manually Corrected Program}
        \label{lst:taco_fixed_code}
    \end{minipage}
    \caption{A critical miscompilation detected by \name for the kernel $A(j) = B(i,j) \cdot C(j)$. Listing \ref{lst:taco_buggy_code} (Left) shows the original TACO-generated code containing an initialization error in the sparse iteration loop. Listing \ref{lst:taco_fixed_code} (Right) shows the manually corrected implementation, highlighting the specific control flow logic required for correct execution.}
    \label{fig:full_comparison}
\end{figure*}

%% file: sections/motivation.tex
\section{Motivation} \label{sec:motivation}

\input{algorithms/taco-dense-kernel}

\begin{equation}
\label{eq:kernel_example}
\underbrace{%
\begin{bmatrix}
9 \\ 16 \\ 0
\end{bmatrix}}_{\mathbf{A}}
=
\underbrace{%
\begin{bmatrix}
0 & 4 & 0 \\
2 & 8 & 0 \\
1 & 0 & 0
\end{bmatrix}}_{\mathbf{B}}
\;\times\;
\underbrace{%
\begin{bmatrix}
0 \\ 2 \\ 5
\end{bmatrix}}_{\mathbf{C}}
=
\begin{bmatrix}
0 \cdot 0 + 2 \cdot 2  + 1 \cdot 5 \\
4 \cdot 0 + 8 \cdot 2 + 0 \cdot 5 \\
0 \cdot 0 + 0 \cdot 2 + 0 \cdot 5 \\
\end{bmatrix}
\end{equation}

\input{algorithms/bug-dense-loop}
\input{algorithms/bug-sparse-loop}

To illustrate the challenges of sparse tensor compilation, consider the kernel: $A(j) = B(i,j) \cdot C(j)$. 
Here, each column of the tensor $B$ is scaled and reduced by the elements of vector $C$ to compute the elements of $A$.
Algorithm \ref{alg:bug-dense-kernel} depicts the standard lowering for dense tensors, where elements are accessed via direct addressing.
In contrast, Algorithm \ref{alg:bug-csr-kernel} demonstrates the corresponding pseudocode when $B$ is stored in CSR format.
The transition to sparse storage forces the compiler to replace affine loop nests with complex data-dependent irregular iterations that traverse compressed index arrays.
Regardless of the underlying format, the computation is expected to produce the deterministic results shown in Equation~\ref{eq:kernel_example}.

To demonstrate the impact of output storage formats on code generation, Listing~\ref{lst:taco_buggy_code} and Listing~\ref{lst:taco_dense_kernel} present the code generated by TACO for this kernel using different storage formats for the output tensor $A$.
While the dense output variant executes correctly, the sparse output variant produces erroneous results.
The defect, exposed by \name, stems from a miscompilation in the loop initialization.
As identified in Listing~\ref{lst:taco_buggy_code} (Lines 2 and 7), the compiler fails to reset the coordinate tracker $j\_A$ within the loop correctly.
The corrected implementation, shown in Listing~\ref{lst:taco_fixed_code}, moves the initialization inside the loop body. 
This example underscores the inherent fragility of STCs: even elementary kernels can trigger errors that evade manual inspection.

Addressing these failures requires automated metamorphic testing.
In the domain of dense tensor compilers, prior work has relied on {\em Computation Graph-based Fuzzing}~\cite{polyjuice, nnsmith}. 
This methodology constructs extensive sequences of tensor operations and mutates the graph topology and individual operations to generate semantic equivalents.
However, this approach is ill-suited for STCs for three reasons.
First, the overhead of graph management incurs prohibitive latency, often limiting throughput to approximately four mutant executions per second~\cite{polyjuice}, thereby restricting test coverage.
Second, graph-level mutations produce bloated failure cases that require computationally expensive minimization to isolate the root cause.
Finally, these tools are constrained to a fixed library of high-level operators, limiting their ability to stress the arbitrary iteration patterns defined by general einsum notation.

Moreover, a wider class of STCs treats tensor operations as isolated lowering targets. 
The compilation process resembles a macro expansion where each einsum expression is translated into a standalone loop nest, oblivious to the broader dataflow context. 
This design characteristic renders graph-based generation ineffective for metamorphic testing. 
Because modifications to one operation in a sequence do not influence the lowering strategy of its neighbors.
Therefore, the search space for bugs is effectively partitioned by operation. 
Hence, a fuzzer that generates long chains of independent operations merely retests the same isolated expansion mechanisms repeatedly, validating the redundancy of graph-level mutation in this domain.

A potential alternative is to employ generic grammar-based fuzzers, such as Grammarinator~\cite{grammarinator}, to synthesize einsum expressions directly. 
However, the validity of the einsum notation is governed by {\em context-sensitive} constraints. 
Standard grammar-based fuzzers, which typically operate on Context-Free Grammars, lack the semantic awareness to enforce these cross-reference constraints.
Consequently, they produce a high volume of invalid kernels that are rejected by the compiler frontend and fail to reach the critical loop-lowering passes.

These limitations motivate the need for a specialized testing framework. 
Effective validation of STCs requires a system capable of (1) generating valid einsum expressions, (2) systematically mutating the expressions to generate semantically equivalent programs, and (3) leveraging metamorphic relations to detect semantic divergence without a reference compiler.

%% file: algorithms/taco-dense-kernel.tex
\begin{listing}[ht]
    %\vspace{-2em}
    \caption{Correct TACO-Generated Program for the Dense Case. Implements the kernel $A(j) = B(i, j) \cdot C(i)$ where $A$ is a dense vector, $B$ is a CSR matrix and $C$ is a sparse vector.}
    \label{lst:taco_dense_kernel}
    
    % FIX: xleftmargin reserves space for numbers so they don't bleed out
    % FIX: linenos=true ensures numbers are actually on (if not set globally)
    \begin{minted}[
        fontsize=\scriptsize, 
        linenos=true, 
        xleftmargin=15pt, 
        numbersep=5pt
    ]{c}
int main() {
  for (int iC = C1_pos[0]; iC < C1_pos[1]; iC++) {
    int i = C1_crd[iC];
    for (int j = 0; j < B2_dimension; j++) {
      int jB = i * B2_dimension + j;
      A_vals[j] = A_vals[j] + B_vals[jB] * C_vals[iC];
    }
  }
}
    \end{minted}
\end{listing}

%% file: algorithms/bug-dense-loop.tex
\begin{algorithm}[t]
\caption{Dense Tensor Computation\\$A(j) = B(i,j) \cdot C(j)$}
\label{alg:bug-dense-kernel}
\DontPrintSemicolon
\SetKwInOut{Input}{Input}
\SetKwInOut{Output}{Output}

\Input{Dense tensor $B \in \mathbb{R}^{m \times n}$, vector $C \in \mathbb{R}^{n}$}
\Output{Vector $A \in \mathbb{R}^{n}$ such that $A_j = B_{ij} \cdot C_j$}

\BlankLine
\For{$j \gets 0$ \KwTo $n-1$}{
    $A[j] \gets 0$\;
    \For{$i \gets 0$ \KwTo $m-1$}{
        $A[j] \gets A[j] + B[i][j] \cdot C[j]$
    }
}
\Return $A$
\end{algorithm}

%% file: algorithms/bug-sparse-loop.tex
\begin{algorithm}[t]
\caption{Sparse Tensor Computation (CSR)\\$A(j) = B(i,j) \cdot C(j)$}
\label{alg:bug-csr-kernel}
\DontPrintSemicolon
\SetKwInOut{Input}{Input}
\SetKwInOut{Output}{Output}

\Input{
Sparse tensor $B$ in CSR format with arrays: 
\begin{itemize}
  \item $B.\text{val}$: nonzero values,
  \item $B.\text{col\_idx}$: column indices,
  \item $B.\text{row\_ptr}$: row pointers; 
\end{itemize}
Vector $C \in \mathbb{R}^{n}$
}
\Output{Vector $A \in \mathbb{R}^{n}$ such that $A_j = \sum_i B_{ij} \cdot C_j$}

\BlankLine
Initialize $A[j] \gets 0$ for all $j = 0,1 \dots n-1$\;

\For{$i \gets 0$ \KwTo $m-1$}{
    \For{$k \gets B.\text{row\_ptr}[i]$ \KwTo $B.\text{row\_ptr}[i+1]$}{
        $j \gets B.\text{col\_idx}[k]$\;
        $A[j] \gets A[j] + B.\text{val}[k] \cdot C[j]$\;
    }
}
\Return $A$
\end{algorithm}

%% file: sections/design_implementation.tex
\section{Design \& Implementation} \label{sec:design}

Unlike dense tensor compilers, which construct a global computation graph to manage lowering and optimization across a sequence of operations, STCs take a different approach.
Rather than maintaining a graph-like structure for the entire operation sequence, STCs focus on constructing an \textit{iteration graph} for a single einsum expression in isolation. 
Hence, the complexity of STCs lies in synthesizing the loop nests for individual operations, rather than performing graph-level rewrites on a multi-node sequence.
In this context, chaining multiple sequences of einsum operations yields diminishing returns for testing, since the complexity lies in the synthesis of iterations rather than in the graph topology. 
Therefore, our fuzzer diverges from graph-based generation; instead, it synthesizes a single einsum notation to represent complex tensor operations, as detailed in Algorithm~\ref{alg:random-einsum}.

\subsection{Random Kernel Generation}
\input{algorithms/random_kernel_generation}

While einsum equations appear syntactically simple, their validity relies on context-sensitive constraints that exceed the expressive power of CFGs. 
A fundamental validity rule is that the set of output indices $\mathcal{O}$ must be a subset of the union of all input indices $\bigcup \mathcal{I}_{\text{in}}$ (\ie $\mathcal{O} \subseteq \bigcup \mathcal{I}_{\text{in}}$).
Therefore, it implies that the validity of the output symbols is dependent on the specific symbols consumed earlier in the input sequence. 
This dependency violates the {\em Pumping Lemma} for Context-Free Languages~\cite{pumping_lemma}, classifying valid einsum expressions as a Context-Sensitive Language.

To synthesize a structurally valid einsum expression, we view the generation process as a constraint satisfaction problem involving \textit{index connectivity} and \textit{dimensional consistency}. 
First, we populate the index sets for all input tensors $(B_1, B_2, \dots, B_N)$ by sampling from a global index pool $\mathcal{I}$, strictly enforcing a maximum rank $R_{\max}$ per tensor. 
We then partition the used indices into two sets: the output indices $\mathcal{O}$ (preserved dimensions) and the contraction indices $S = \mathcal{I} \setminus \mathcal{O}$. 
A critical validity constraint in einsum semantics is that a contraction index typically bridges dimensions across tensors. 
If an index $s \in S$ appears in fewer than two input tensors, the contraction is ill-defined. 
To resolve this, we enforce connectivity: we iterate through $S$ and inject any under-represented contraction indices into an additional randomly selected input tensor. 
Once the symbolic structure is validated, we map the abstract indices to concrete runtime dimensions. 
By assigning a distinct integer size to each unique index in $I$ and propagating these sizes to the corresponding axes of $(B_1, B_2, \dots, B_N)$, we guarantee that the generated tensors possess compatible shapes for the specified contraction.

\subsection{Mutation Operators}

However, constraining the search space to individual einsum operations precludes using traditional graph-level rewrite rules as mutation operators. 
We address this by introducing two invariance-based mutation strategies: \textit{Algebraic Commutativity} and \textit{Storage Format Heterogeneity}.

The algebraic commutativity is grounded in tensor algebra: assuming the tensor elements belong to a commutative ring (\eg the field of real numbers $\mathbb{R}$), the multiplication operation is commutative (\ie $\forall\, a,b \in \mathbb{R},\; a \cdot b = b \cdot a$). 
This property extends to einsum operations, where the order of operands does not alter the semantic result. 
For instance, the expression $A({i,j}) = B({i,k}) * C({k,j})$ is semantically equivalent to $A({i,j}) = C({k,j}) * B({i,k})$. 
By permuting operands, we force the compiler to generate different iteration schedules for the same mathematical operation.

The second operator, \textit{Storage Format Heterogeneity}, targets the complexity of sparse iteration schemes for different compressed storage formats. 
In sparse tensor compilation, the mathematical definition of a kernel is orthogonal to the physical layout of its data.
A tensor $B(i,j)$ contains the same nonzero values in the exact coordinates whether stored in COO, CSR, CSC, or any other compressed storage format.
However, the choice of format significantly alters the code-generation path, as shown in Figure \ref{fig:full_comparison}.
We leverage this decoupling by randomly assigning distinct storage formats to each input and output tensor in the generated einsum expression. 
This forces the compiler to synthesize unique iteration graphs and loop nests for every format combination. 
Because the semantics of the computation remain invariant, any divergence in the output across these format permutations indicates a compilation fault in handling specific access patterns or compressed data formats.
By composing these two mutation operators\textemdash permuting operand order and varying storage formats\textemdash we create a rich space of semantically equivalent test programs from a single einsum expression.

\input{figures/main_fuzz_cycle}

\subsection{Language-Agnostic Architecture}

While our mutation operators generate a rich space of abstract test programs, executing them requires navigating the syntactic fragmentation of the STC landscape.
%\km{I'm not sure "macro expanders" is the right word to use here.}
Since most STCs operate as macro expanders within high-level host languages (\eg C++ for TACO, Julia for Finch), a direct-generation approach is not inherently portable. 
To ensure broad compatibility, we designed a generic JSON abstraction layer that captures all semantic details of the synthesized kernel, including randomized storage-format configurations. 
This design shifts the integration burden from the fuzzer to a thin abstraction layer: compiler developers can integrate their tools simply by implementing a lightweight translator that parses this JSON schema and emits the corresponding domain-specific sparse tensor program.

Figure~\ref{fig:main_fuzz} illustrates the decoupled execution workflow of our fuzzer. 
The core fuzzing engine is isolated from the target STC and interacts solely through the abstraction layer. 
At runtime, the fuzzer first dispatches a randomly selected reference program to the target STC via the translator. 
Upon successful execution, it generates semantically equivalent mutants\textemdash applying the commutativity and storage format heterogeneity operators\textemdash and executes them against the same backend. 
The fuzzer then acts as a metamorphic oracle, comparing the tensor outputs of the mutants against the reference result. 
Any discrepancy flags a correctness violation in the STC's iteration schemes. 
This plugin-based architecture ensures the system remains extensible to future research and industrial compilers.

\subsection{Program Execution \& Bug Detection}

Executing a tensor kernel requires not only the code but also valid input data. 
Complementing its program generation, \name automatically synthesizes small input tensors that strictly adhere to the \textit{dimensional consistency} of the einsum specification.
Once the reference program successfully completes execution, its output is recorded as the ground truth. 
The mutated program is then executed, and its results are compared against this baseline. 
If a discrepancy is detected, \name aggregates the input tensors, both program variants, and their respective outputs into a detailed bug report.

%% file: algorithms/random_kernel_generation.tex
\SetKwInput{Step}{Step}

\begin{algorithm}[t]
\caption{Random Einsum Expression Generation}
\label{alg:random-einsum}
\DontPrintSemicolon
\SetKwInOut{Input}{Input}
\SetKwInOut{Output}{Output}

\Input{
$N$: number of input tensors,\\
$R_{\max}$: maximum tensor rank,\\
$I$: pool of index labels
}
\Output{
Random einsum expression $A = B_1 * B_2 * \dots * B_N$
}

\BlankLine
\Step{1: Assign indices to input tensors}
\For{$k \gets 1$ \KwTo $N$}{
    $r_k \gets \text{UniformRandom}(1, R_{\max})$\;
    $B_k.\text{indices} \gets$ randomly select $r_k$ distinct indices from $\mathcal{I}$\;
    Update usage count of selected indices\;
}

\Step{2: Determine output tensor indices}
$\mathcal{O} \gets$ random subset of indices with non-zero usage\;

\Step{3: Ensure valid reduction indices}
\For{each index $i \in \mathcal{I} \setminus \mathcal{O}$ with usage count $= 1$}{
    Add $i$ to a different input tensor to ensure it occurs at least twice\;
}

\Step{4: Assign tensor shapes}
\For{each index $i \in \mathcal{I}$}{
    assign random dimension $d_i$\;
}
\For{each tensor $B_k$}{
    \For{each $j$ in $B_k.\text{indices}$}{
        $B_k.\text{shape}[j] \gets d_{B_k.\text{indices}[j]}$\;
    }
}

\Step{5: Construct einsum expression}
$A(\mathcal{O}) = \prod_{k=1}^{N} B_k(B_k.\text{indices})$\;

\Return $A$ and $\{B_1, \dots, B_N\}$
\end{algorithm}

%% file: figures/main_fuzz_cycle.tex
\begin{figure*}
    \centering
    \includegraphics[width=1\linewidth]{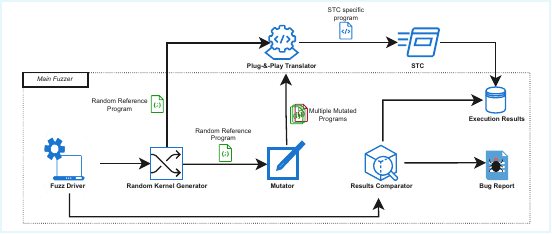}
    \caption{The main fuzzing loop. The system generates a random tensor program and a few mutated variants. Both versions are compiled and executed, and the outputs are compared to identify mismatches indicating compiler bugs.}
    \label{fig:main_fuzz}
\end{figure*}

%% file: sections/evaluation.tex
\section{Evaluation} \label{sec:evaluation}

To empirically quantify the advantage of our generation strategy, we established a baseline using \textit{Grammarinator}~\cite{grammarinator}, a state-of-the-art grammar-based fuzzer configured with a standard context-free grammar for einsum notation. 
We conducted a large-scale comparative study, synthesizing a corpus of one million ($10^6$) kernels using both the baseline and our custom generator.

The results reveal a stark disparity in the efficiency of the generation. 
While our constraint-based algorithm guarantees a 100\% validity rate by construction, the grammar-based baseline produced valid kernels in only $\sim$3.3\% of attempts. 
The vast majority of grammar-generated inputs were rejected due to dimensional inconsistencies. This result confirms that generic grammar fuzzers are fundamentally inefficient for tensor compiler testing, as they waste over $\sim$96\% of the fuzzing budget on syntactically valid but semantically malformed inputs.

We integrated our constraint-based einsum generator into the \name framework to serve as the core input synthesis engine. 
To evaluate its effectiveness, we conducted a sustained six-hour fuzzing campaign targeting two state-of-the-art sparse tensor compilers: TACO and Finch, and the results can be found in Table \ref{tab:comparison}.

During the fuzzing lifecycle, \name categorizes execution anomalies into three distinct failure scenarios that vary in diagnostic fidelity. 
The first scenario involves the failure of the initial reference kernel; since many STCs support only a strict subset of the einsum standard, these rejections frequently stem from unsupported features rather than genuine defects.
The second scenario occurs when the reference kernel executes successfully, but a semantically equivalent mutant triggers a compilation error or runtime crash.
While this often indicates a valid crash bug, it may also stem from gaps in the compiler's support for specific iteration schemes.
The third and most critical scenario arises when both the reference and mutant programs execute successfully but yield divergent outputs. This constitutes a \textit{functional correctness violation}, or silent miscompilation.
Unlike crashes, which are often caught by runtime assertions, these divergences indicate that the compiler has generated incorrect sparse iteration schemes for a valid mathematical operation.

\begin{table}[t]
    \centering
    \caption{Bugs found by \name over a 6-hour fuzzing campaign.}
    \label{tab:comparison}
    
    \begin{tabular}{l cccc}
        \toprule
        
        \textbf{STCs}
        
         & \#Iterations & STC-NA & C-Bugs & WC-Bugs \\
        \midrule
        
        % TACO & 267,231 & 236,235 & 14,441 & 5,758 \\
        TACO & 267k & 236k & 14.4k & 5,758 \\
        
        Finch & 1,619 & 7 & 57 & 0 \\
        
        \bottomrule
    \end{tabular}

    % TABLE NOTE for Definitions
    \begin{minipage}{0.48\textwidth} % Adjust width to match table
        \vspace{5pt}
        \footnotesize
        \textbf{Note:} STC-NA: Not Acceptable Input for STC; C-Bugs: Number of Crash Bugs; WC-Bugs: Number of Wrong-Code Bugs (Silent Errors). 
        '\#Iterations' denotes total fuzzing iterations.
    \end{minipage}
    \vspace{-1em}
\end{table}

\subsection{TACO Evaluation Results} 
\label{sec:taco_eval}

Among the valid einsum operations accepted by the compiler's frontend, \name exposed defects in approximately $\sim$65.2\% of cases. Crucially, $\sim$18.6\% of these failures were classified as critical miscompilations (wrong code bugs), where the compiler silently generated incorrect code.

We attribute this high volume of defects to two primary factors. 
First, the reported figures reflect total failure counts; determining unique root causes would require exhaustive manual inspection, which was infeasible given the scale of failures. 
Second, TACO was designed primarily as a foundational research prototype to demonstrate sparse compilation concepts, rather than as a production-hardened system. 
As the project is no longer actively maintained or accepting bug reports, we focused our analysis on aggregate failure rates to demonstrate the fuzzer's efficacy, rather than performing granular deduplication for reporting purposes.
Due to the black-box nature of \name, reliable deduplication is inherently difficult, as a single underlying compiler defect can be triggered by multiple, syntactically distinct einsum expressions.

\subsection{Finch Evaluation Results}
\label{sec:finch_eval}

While our campaign against TACO evaluated over 267,000 kernels and uncovered thousands of defects, the evaluation on Finch was constrained by significant compilation latency, completing only 1,619 iterations in the same six-hour window.
Despite this reduced throughput, \name successfully identified 57 crash bugs in the Finch compiler. No silent miscompilations were detected within this limited sample size, though the presence of crash-inducing inputs confirms the fuzzer's ability to stress the Julia-based infrastructure.

Two factors explain this disparity. 
First and most significantly, the compilation time for Finch kernels is the primary bottleneck. 
As shown in Figure~\ref{fig:runtime_comparison}, TACO processed a single kernel in approximately $64$ ms, whereas Finch required over $181$ s per kernel. 
The complicated data layout of sparse tensors adds significant overhead to the \textit{Initialization} and \textit{Computation} compilation stages, a cost that is incurred repeatedly for every iteration.
Second, the current evaluation was restricted to a preliminary six-hour window; we reserve more extensive, longitudinal fuzzing campaigns for future work.

One way to improve throughput is to reduce the granularity of the test variations. By isolating changes to specific expressions, we could allow the Julia compiler to only recompile the mutated parts instead of the whole program. However, this requires structural changes and specialization of the fuzzer.

\input{figures/taco_finch_eval}

%% file: figures/taco_finch_eval.tex
\begin{figure}[t]   
	\vspace{-2em}  
	\centering
    \includegraphics[width=\linewidth]{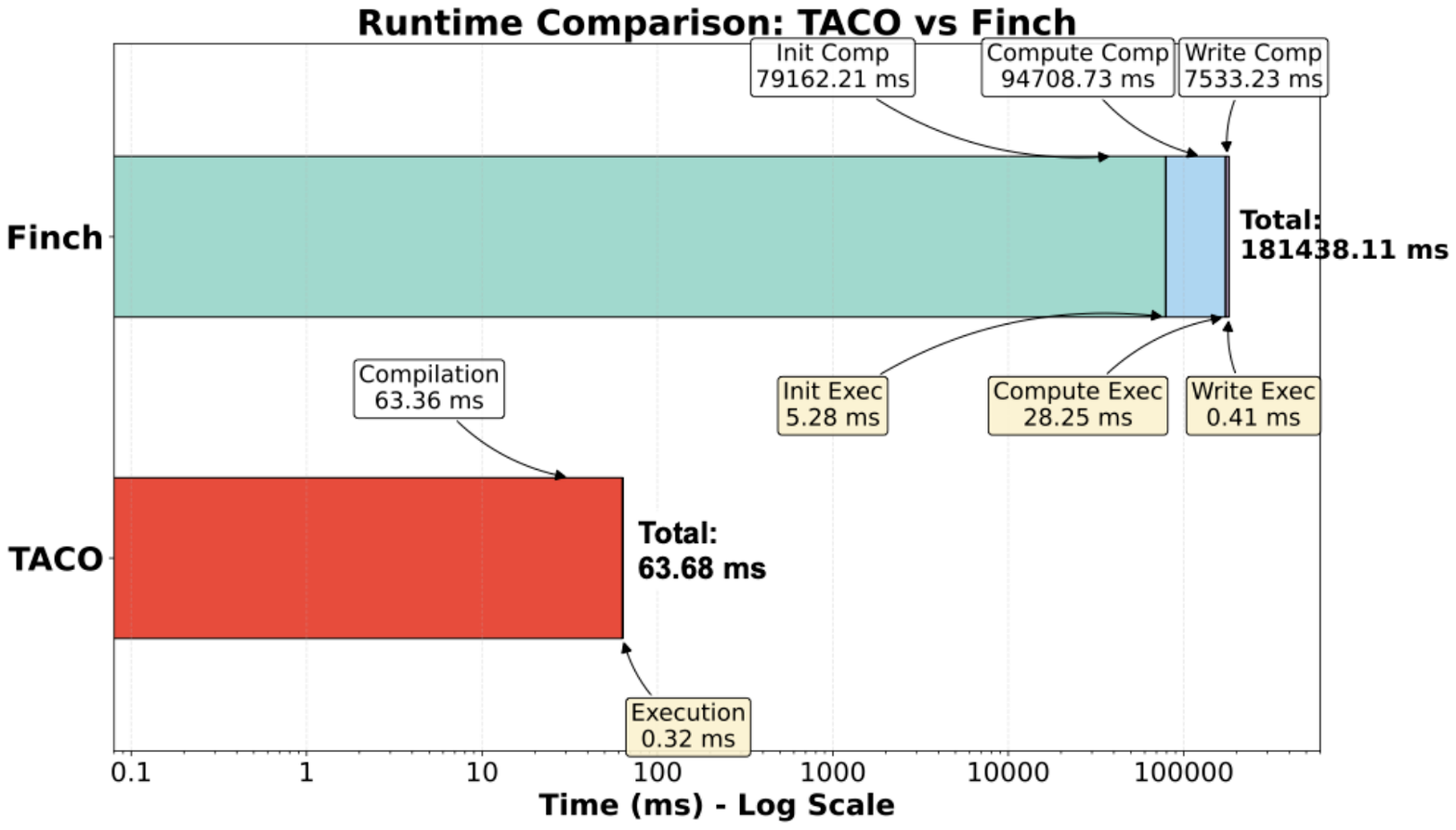}
    \vspace{-2em}
    \caption{Runtime Comparison: TACO vs Finch. The time axis uses a log scale to capture the disparity in compilation times.}
    \label{fig:runtime_comparison}
\end{figure}

%% file: sections/discussion.tex
\section{Discussion and future work} \label{sec:discussion}

A persistent challenge in validating numerical software is the handling of floating-point determinism. 
While our mutation operator exploits the commutative property of einsum operations, this high-level reordering often compels the compiler to alter the low-level reduction schedule. 
Since IEEE-754 floating-point arithmetic is not associative, these scheduling changes introduce minor numerical deviations. 
Currently, \name employs a relaxed $\epsilon$-based comparator to tolerate this noise; however, this introduces a trade-off in which loose tolerances may mask subtle correctness bugs.
In the future, we will mitigate this by introducing integer-only test modes to validate synthesized control flow strictly. 
This approach eliminates numerical noise, enabling the unambiguous distinction between valid algorithmic reordering and genuine miscompilation.

Beyond addressing precision challenges, a promising approach for enhancing the fuzzer is the systematic exploitation of tensor transposition. 
Currently, our strategy relies on operand commutativity and storage format heterogeneity.
However, by leveraging the algebraic invariance of tensor contractions under index permutation, we can introduce a new class of Transpose Mutation Operators. 
These operators would transform the input tensors by permuting their dimensions (effectively computing $A^T$) while adjusting the contraction indices to preserve semantic equivalence. 
This targeted approach would rigorously stress-test the compiler's ability to handle permuted access patterns, exposing defects in the index map generation logic that simple operand reordering cannot reach.

% We recognize that advanced STCs often expose unique features, such as specific hardware tiling directives or loop fusion pragmas, that offer rich targets for fuzzing but are not generalizable to the broader ecosystem. 
% To support these without compromising the framework's language-agnostic core, we plan to design an extensible mutation interface. 
% We envision a system in which users implementing an STC-specific translator can define bespoke mutation operators that integrate directly into the generation pipeline. 
% This would allow the fuzzer to exploit specialized compiler features while maintaining a clean separation of concerns.

Moreover, we do not employ advanced deduplication strategies for redundant bug reports (Section \ref{sec:taco_eval}), nor do we implement compiler-specific optimizations to accelerate compilation times for targets like Finch (Section \ref{sec:finch_eval}). 
Addressing these challenges would require inspecting internal compiler states or tailoring the generator to specific backends, modifications that would fundamentally alter \name from a general-purpose black-box fuzzer into a specialized grey- or white-box tool.

In future work, we plan to expand our evaluation by benchmarking \name against a broader range of state-of-the-art baselines~\cite{ISLa, nautilus, fuzz4all} to further validate its efficacy. 
Finally, we aim to demonstrate the extensibility of our framework by integrating emerging platforms such as the MLIR Sparse Dialect and PyTorch Sparse.
While integrating PyTorch Sparse is straightforward due to its native einsum support, targeting MLIR Sparse Dialect presents a significant semantic gap, as the dialect does not directly consume high-level einsum declarations. 
Consequently, this integration will require enhancing our abstraction layer to translate our JSON-based intermediate representation into the MLIR Sparse dialect, thereby enabling a robust comparative analysis of diverse sparse compiler infrastructures.

%% file: sections/related_work.tex
\section{Related Work} \label{sec:related}

Automated compiler validation is a well-established area of research. 
The seminal work of Csmith~\cite{csmith} demonstrated the efficacy of random differential testing, revealing hundreds of bugs in GCC and LLVM by generating valid C programs from scratch. 
Building on this foundation, subsequent frameworks such as Orion~\cite{compiler_emi} and YarpGen~\cite{yarpgen} introduced sophisticated mutation-based strategies. 
These tools create semantically equivalent test cases by injecting dead code or simplifying arithmetic expressions. 

The proliferation of deep learning compilers has catalyzed the development of specialized testing frameworks. 
NNSmith~\cite{nnsmith} pioneered the constraint-based generation of valid computation graphs, excelling at verifying diverse graph topologies and operator combinations.
Building on this foundation, PolyJuice~\cite{polyjuice} employed equality saturation to synthesize semantically equivalent graph variants, specifically targeting defects in graph-level optimizers. 
At the same time, HiraGen~\cite{hiragen} focused on stressing high-level optimization passes through systematic graph mutations.

More recently, Large Language Models (LLMs) have been adapted for input generation. 
TitanFuzz~\cite{titanfuzz} utilizes LLMs in a black-box manner to synthesize operator sequences without internal compiler knowledge. 
In contrast, WhiteFox~\cite{whitefox} introduces a white-box approach that leverages LLMs to analyze the compiler's source code and optimization passes to generate targeted test inputs. 
However, all these frameworks primarily target dense computation graphs constructed from standard operators, leaving the specific challenges of sparse tensor compilation and lowering phases largely unaddressed.

The landscape of sparse compilation is rapidly evolving, driven by the need to decouple algorithm specification from data representation. 
TACO~\cite{taco} pioneered the concept of format-agnostic compilation, introducing a lattice-based theory to synthesize code for arbitrary sparse tensor formats. 
This foundational work has since been adopted by modern frameworks, including MLIR's Sparse Tensor Dialect~\cite{sparse_mlir}, Finch~\cite{Ahrens2025Finch}, and TVM's SparseTIR~\cite{sparse_tir}, as well as industrial extensions to XLA~\cite{tensorflow_sparse} and PyTorch~\cite{scorch, pytorch_sparse}. 
As these systems transition from research prototypes to critical production infrastructure, the need for robust, specialized testing frameworks becomes increasingly acute.

% While STCs share general compilation risks with their dense counterparts, they introduce a distinct class of defects related to compressed storage formats and coordinate-based index traversal. 
% Furthermore, the architectural maturity of STCs differs significantly from dense frameworks. 
% Whereas dense compilers perform aggressive cross-kernel optimization on global computation graphs, current STCs typically function as sequential macro expanders, lowering operations individually rather than performing holistic graph-level analysis. 
% Consequently, fuzzers designed to stress graph optimizers fail to exercise the critical, localized lowering mechanisms that govern sparse tensor iterations. 

Grammar-based fuzzing is a standard technique for testing language processors. 
Tools such as Grammarinator~\cite{grammarinator}, LangFuzz~\cite{langfuzz}, and Nautilus~\cite{nautilus} generate inputs by traversing formal grammars, proving highly effective for syntactic testing across a wide range of languages. 
However, generic grammar-based fuzzers, which typically operate on Context-Free Grammars, lack the intrinsic mechanism to enforce these semantic constraints required by the sparse tensor programs in einsum notation.

%However, directly applying these tools to STCs presents a challenge. 
%Valid einsum expressions are governed by strict context-sensitive dependencies, specifically the requirement that contraction indices must maintain dimensional consistency across multiple tensor operands. 
%Generic grammar fuzzers, which typically operate on Context-Free Grammars, lack the intrinsic mechanism to enforce these semantic constraints. 
% Our work addresses this limitation by supplanting purely grammar-based generation with a constraint-based algorithm, designed to guarantee semantic validity while retaining the expressive power of random input generation.

%% file: sections/conclusion.tex
\section{Conclusion} \label{sec:conclusion}

In this paper, we present \name, the first extensible black-box fuzzing framework specifically architected for the automated testing of STCs.
Unlike prior frameworks constrained by fixed operator libraries~\cite{nnsmith, polyjuice}, \name leverages einsum notation as a fully general input abstraction to synthesize arbitrary tensor contractions.

We demonstrated the utility of \name by targeting two distinct STCs, TACO and Finch.
Our evaluation of TACO revealed significant robustness issues, with the fuzzer exposing crashes or miscompilations over $\sim$60\% of generated test cases.
These findings, alongside the successful adaptation to the Finch ecosystem, underscore the fragility of the current sparse compilation landscape and highlight the necessity of specialized tools like \name to ensure the reliability of next-generation infrastructure.

\newpage